\documentclass[11pt,reqno]{article}
\usepackage{amsmath}
\usepackage{amsfonts}
\usepackage{amssymb}
\usepackage{hyperref}
\usepackage{latexsym}
\usepackage[dvips]{graphicx}
\usepackage{epsf}
\usepackage{color}
\usepackage{bbold}

\allowdisplaybreaks \numberwithin{equation}{section}

\newcommand{\be}{\begin{equation}}
\newcommand{\ee}{\end{equation}}

\newcommand{\f}{\frac}
\newcommand{\p}{\partial}
\newcommand{\na}{\nabla}
\newcommand{\Tr}{{\rm Tr}}
\newcommand{\tr}{{\rm tr}}

\let\a=\alpha \let\b=\beta    \let\d=\delta
\let\z=\zeta     \let\th=\theta   
\let\m=\mu    \let\n=\nu          \let\r=\rho \let\om=\omega
\let\s=\sigma      
    
\let\G=\Gamma \let\D=\Delta   \let\L=\Lambda 
         
  \let\eps=\epsilon

\newcommand{\gb}{\bar{g}}

\newcommand{\Rb}{\bar{R}}

\newcommand{\Gb}{\bar{\Gamma}}

\newcommand{\Cb}{\bar{C}}

\newcommand{\nab}{\bar{\nabla}}

\newcommand{\tG}{\tilde{G}}

\newcommand{\cL}{\mathcal{L}}

\newcommand{\cR}{\mathcal{R}}

\newcommand{\ug}{\underline{g}}

\begin{document}

\thispagestyle{empty}
\bigskip

\begin{center}
 {\LARGE\bfseries  Essential nature of Newton's constant\\ in unimodular gravity}
\\[10mm]
{\large Dario Benedetti}
\\[3mm]
{\small\slshape
Laboratoire de Physique Th\'eorique, CNRS UMR 8627,\\
B\^at. 210, Universit\'e Paris-Sud, 91405 Orsay Cedex, France \\ 
\vspace{.3cm}
 {\upshape\ttfamily dario.benedetti@th.u-psud.fr 
 } }
\end{center}
\vspace{5mm}

\hrule\bigskip
\centerline{\bfseries Abstract} \medskip
%

We point out that in unimodular gravity Newton's constant is an essential coupling, i.e. it is independent of field redefinitions. We illustrate the consequences of this fact by a calculation in a standard simple approximation, showing that in this case the renormalization group flow of  Newton's constant is gauge and parametrization independent.

\bigskip
\hrule\bigskip
\tableofcontents

\section{Introduction}

Unimodular gravity is a theory of gravitation in which the metric determinant is a fixed non-dynamical density.
Despite being almost as old as general relativity \cite{Einstein:1919gv}, and classically equivalent to it, unimodular gravity has never reached the popularity of the standard formulation.  Nevertheless, it is from time to time retrieved by different authors (e.g. \cite{vanderBij:1981ym,Unruh:1988in,Henneaux:1989zc,Alvarez:2005iy,Smolin:2009ti,Ellis:2010uc,Saltas:2014cta,Alvarez:2015pla}) for a number of attractive features. Most notably, in unimodular gravity the cosmological constant appears as an integration constant rather than as a coupling in the action, and therefore it is not subject to quantum corrections. 
Such feature is of course attractive in the context of the cosmological constant problem, although by itself it is not a full solution \cite{Smolin:2009ti}. The unimodularity constraint has also been argued to solve the problem of time \cite{Unruh:1988in}, but that is also not free from difficulties \cite{Kuchar:1991xd}.
At a more technical level, it can help us in several ways, for example, by recasting the action in polynomial form \cite{vanderBij:1981ym}, or by getting rid of some ambiguities in the path integral measure (the DeWitt supermetric is trivially independent of the C-ambiguity).

Ultimately, despite the several advantages that unimodular gravity (UG) has compared to general relativity (GR), the main big problems remain open also in this formulation, and in particular as a quantum theory UG is equally non-renormalizable as GR. Whether these two formulations of the classical theory both admit a UV completion in the quantum domain, and whether these would be equivalent as well, is an open question, and it will probably be so as long the challenge of quantizing gravity remains open.
Preliminary calculations by Eichhorn \cite{Eichhorn:2013xr,Eichhorn:2015bna} suggest that unimodular gravity might have a UV completion in the form of an asymptotic safety scenario \cite{Weinberg:1980gg,Niedermaier:2006wt}, just like more extensive calculations have indicated for non-unimodular gravity (for a review up to 2012 see \cite{Reuter:2012id}, for more recent results see \cite{Falls:2014tra,Demmel:2015oqa,Christiansen:2015rva} and references therein). 
In fact, from the point of view of a standard field theoretic quantization, the unimodularity constraint might be seen just as a (partial) gauge-fixing of the non-unimodular theory, as done recently in \cite{Percacci:2015wwa}, and therefore one might expect equivalence between the two formulations. However, one should notice that when implementing the unimodularity condition as a gauge-fixing, the usual Faddev-Popov ghosts need to be included as well, whereas they are not needed if the restriction is part of the fundamental definition of the theory.
We will discuss this point more explicitly in the following.

In this note, we want to stress some features of unimodular gravity that make it appealing from the point of view of the renormalization group.
First of all, one should notice that because of the unimodularity condition, the scale factor is not being integrated over in the gravitational path integral, it is fixed and identical to the background one. Therefore, talking about scale dependence and renormalization group flow sounds less disturbing: the theory is not completely background independent, there is some fixed background structure that provides a scale.
For the same reason, it is not possible in UG to absorb Newton's constant by a metric redefinition, because such a redefinition (a simple rescaling in fact) would change the determinant and violate the unimodularity condition. Therefore, we expect $G$ to be an essential (i.e. non-redundant \cite{Wegner:1976nq,Weinberg:1980gg}) coupling, unlike in the non-unimodular case \cite{Weinberg:1980gg,Percacci:2004sb,Benedetti:2011ct}.
This fact will allow us to extract the beta function of Newton's coupling (in a simple truncation of the functional renormalization group equation) using an on-shell background, without the complications encountered in \cite{Benedetti:2011ct}. Remembering that only the on-shell effective action is gauge and parametrization independent \cite{Buchbinder:1992rb}, one can understand that since in the non-unimodular case we can only read off the running of Newtons constant off-shell, its beta function is generically gauge and parametrization dependent.
On the contrary, in the unimmodular case we will obtain a gauge and parametrization independent beta function. We should stress this point by noticing the following. In \cite{Falls:2015qga} a gauge-independent beta function for Newton's constant in the non-unimodular case was obtained by choosing a specific parametrization. Although that is a nice result, we find it not completely satisfactory: the general beta function depends both on gauge and parametrization, therefore one can expect to be able to choose one in order to cancel the other, but this still leaves us with an unwanted dependence on such choice.
We will see here that in the unimodular case, due to the intrinsic restriction on the configuration space, we can obtain complete independence on both gauge and parametrization.

In order to illustrate our point, we will work within the functional renormalization group (FRG) framework (e.g. \cite{Gies:2006wv,Rosten:2010vm}), truncating the effective action to the simplest approximation, retaining only Newton's constant. Technically, our calculation is very close to the one in \cite{Eichhorn:2013xr}, but we will work with a general gauge, general parametrization, and in general dimension. 
The calculation is presented in Sec.~\ref{conto}, broken down into several subsections in order to facilitate a quick access to it. We will summarize and comment further on our findings in Sec.~\ref{concl}

\section{RG flow in the unimodular Einstein-Hilbert truncation}
\label{conto}

\subsection{Classical action  and essential nature of Newton's constant}
\label{essential}

The classical action for unimodular gravity is the usual  Einstein-Hilbert action,
\be \label{Sclass}
S[g] = -\f{1}{16\pi G} \int d^d x\, \om\, R \, ,
\ee
but as anticipated, because the volume element is fixed to 
\be \label{unim}
\sqrt{g}=\om\, ,
\ee
where $\om$ is a fixed density,
the action is independent of the cosmological constant, which only appears as a constant of integration of the field equations.
In the pure gravity case, the latter read\footnote{In order to do the functional variation, we can either remember that the linear perturbation must be traceless, and thus straightforwardly obtain the traceless equations \eqref{eom}, or we can include the unimodular constraint in the action by means of a Lagrange multiplier and then eliminate the latter from the equations of motion, obtaining of course the same result.}
\be \label{eom}
R_{\m\n} - \f{1}{d} g_{\m\n} R = 0\, .
\ee
The cosmological constant appears by taking the divergence of the field equations, and using the Bianchi identities, leading to $\p_\m R =0$. Therefore, $R = \bar\Lambda$, for some constant $\bar\Lambda$ (the usual definition of the cosmological constant $\Lambda$ is obtained by setting $\bar\Lambda = \f{2d}{d-2}\Lambda$), and \eqref{eom} can be rewritten as the Einstein equations with a cosmological constant.

The equations of motion play an important role both in the classical and the quantum theory. 
If we redefine $g_{\m\n}\to g'_{\m\n}[g]$, the classical action transforms as $S[g]\to S[g']\simeq S[g] + (\d S[g]/\d g_{\m\n}) (g'_{\m\n}[g]-g_{\m\n})$, and therefore we can eliminate a term from $S[g]$ only if it is proportional to $\d S[g]/\d g_{\m\n}$, that is, the equations of motion.
This trivial statement translates at the quantum level in a similar statement for the generating functional of one particle irreducible diagrams, i.e. the effective action. As a consequence, one can prove gauge and parametrization independence only for those terms in the effective action that are not zero on-shell \cite{Buchbinder:1992rb}.
The same conclusion carries over to the renormalization group flow, where one defines a redundant \cite{Wegner:1976nq} (or inessential \cite{Weinberg:1980gg}) operator to be an eigenoperator of a fixed point which is proportional to the equations of motion of the latter (see also \cite{Dietz:2013sba} for a recent review of this concept).

Now we can make our key observation: by construction, the field equations have vanishing trace, and therefore the action \eqref{Sclass} is not proportional to them. We conclude that in UG Newton's constant is not redundant, i.e. it is an essential coupling. This should be contrasted to the case of GR, where $R$ is proportional to the trace of the Einstein tensor, which in that case is the left-hand-side of the field equations. From a practical point of view, this means that while in GR if we go on-shell we are not able to distinguish anymore the two couplings of the Einstein-Hilbert action $S_{EH} = \f{1}{16\pi G} \int d^d x\, \sqrt{g}\, (2\L-R) $, in UG we can go on-shell and still be able to track the only coupling present in the action \eqref{Sclass}.

 The essential nature of $G$ in UG is the main observation of this paper, and as anticipated in the introduction, it can be simply understood as the fact that we cannot absorb $G$ by a rescaling of the metric, because rescalings are not allowed by the unimodularity constraint.

From the point of view of the coupling the situation might be compared to that of the linear $O(N)$ model versus  the $O(N)$  non-linear sigma model. Consider the action $S=Z \int d^d x\, \p_\m \phi_a \p^\m \phi^a$ with $a=1\ldots N$. In the linear $O(N)$ model, the field $\phi_a$ does not satisfy any constraint, therefore we can absorb $Z$ by a rescaling. In other words, $Z$ is the usual wave function renormalization, which is a redundant coupling.
In the $O(N)$ non-linear sigma model instead, the fields satisfy $\phi_a  \phi^a=1$, and therefore they cannot be rescaled. In such case, $Z$ is an essential coupling, for which it makes sense to look for a fixed point (and in fact it is well known that it has a non-trivial one for small $\eps=d-2$).

\subsection{Parametrization of the metric perturbations}

The natural parametrization for the metric perturbations in unimodular gravity is
\be \label{exp-par}
g_{\m\n} = \gb_{\m\r} (e^{h})^\r_\n
\ee
where  $\sqrt{\gb}=\om$,
\be
(e^{h})^\r_\n = \d^\r_\n + h^\r_\n + \f12 h^\r_\s  h^\s_\n + \ldots \, ,
\ee
and $h^\r_\n = \gb^{\r\m}h_{\m\n}$ for a symmetric tensor $h_{\m\n}$, required to be traceless, i.e. $h^\m_\m=0$.
As a consequence of the tracelessness of the fluctuation field, the determinant of the full metric coincides with the one of the background metric, $g=\gb=\om^2$. 

Although \eqref{exp-par} is the most natural parametrization, it is not unique. We can expand the metric as
\be \label{level-pert}
g_{\m\n} = \gb_{\m\n} + \eps h^{(1)}{}_{\m\n} + \f{\eps^2}{2} h^{(2)}{}_{\m\n} + \f{\eps^3}{6} h^{(3)}{}_{\m\n} +\ldots \, ,
\ee
and impose the unimodularity condition $g=\gb$ order by order in $\eps$.
We find
\be \label{unim1}
h^{(1)}{}^\m_\m = 0 \, ,
\ee
\be \label{unim2}
h^{(2)}{}^\m_\m =  h^{(1)}{}_{\m\n} \, h^{(1)}{}^{\m\n} \, ,
\ee
\be \label{unim3}
h^{(3)}{}^\m_\m = - 2 h^{(1)}{}_\m^\n \, h^{(1)}{}_\n^\r \, h^{(1)}{}_\r^\m + 3  h^{(2)}{}_{\m\n}\, h^{(1)}{}^{\m\n} \, ,
\ee
and so on.
If we allow only unltralocal field redefinitions, we can view the expansion \eqref{level-pert} as a field redefinition of the exponential one, where\footnote{Note that because $h_\m^\m=0$ we have only two possible ultralocal terms at order $\eps^2$ instead of the general four discussed in \cite{Gies:2015tca}.}
\be \label{redef}
h_{\m\n}\to \eps h_{\m\n}+\f{\eps^2}{2} \left( a_1 h_\m^\r h_{\r\n} + a_2 \gb_{\m\n} h_{\r\s} h^{\r\s} \right) + \ldots \, ,
\ee
with the identifications
\be
h^{(1)}{}_{\m\n} = h_{\m\n} \, ,
\ee
\be
h^{(2)}{}_{\m\n} = (1+a_1) h_\m^\r h_{\r\n} + a_2 \gb_{\m\n} h_{\r\s} h^{\r\s} \, ,
\ee
etc.
In this case, the condition \eqref{unim1} remains $h_\m^\m=0$, while \eqref{unim2} enforces $a_1+d a_2=0$.

\subsection{Truncation of the effective average action}

As anticipated, we will work within the functional renormalization group (FRG) framework (e.g. \cite{Gies:2006wv,Rosten:2010vm}), which has first been applied to gravity in  \cite{Reuter:1996cp}.
The central object of the FRG is the average effective action \cite{Wetterich:1992yh,Morris:1993qb} which reduces to the usual effective action in the limit in which the cutoff is removed (i.e. cutoff scale $k$ goes to zero). The most common non-perturbative approximation thus consists in truncating the theory space in which the effective average action is defined.

In the following we are going to derive the beta function for $G$ in the FRG framework, using the crudest, but most pedagogical, truncation.
Following  \cite{Reuter:1996cp}, it is useful to cast a general truncation of the effective average action into the form
\be
\G_k[\bar{\Phi},\Phi] = \Gb_k[g] + \widehat{\G}_k[h, \gb] + 
\G_{\rm gf}[h, \gb] + \G_{\rm gh}[h, \gb, {\rm ghosts}] + S_{\rm aux}[\gb, {\rm aux.fields}] \, .
\ee
 In this decomposition $\bar{\Gamma}_k[\ug]$ depends only on the total metric. $\G_{\rm gf}$ and $\G_{\rm gh}$ denote the gauge-fixing and ghost-terms respectively, for which we will take the classical functionals, eventually allowing a running of their parameters, while $S_{\rm aux}$ is a coupling-independent action encoding the Jacobians that might arise from field redefinitions. $\widehat{\Gamma}_k[h, g]$ encodes the separate dependence on background metric and fluctuations, it vanishes for $h=0$, 
 and it is often thought to capture the quantum corrections to the gauge-fixing term. 
 The role of $\widehat{\Gamma}_k[h, g]$ has been investigated recently in the so-called bimetric truncations (e.g. \cite{Becker:2014qya}) or level expansions (e.g. \cite{Christiansen:2012rx,Codello:2013fpa}); in the present work we will make use of the approximation $\widehat{\Gamma}_k=0$, sometimes referred to as ``single-metric''.

The diffeomorphism invariant part of our truncation is taken to be of the form \eqref{Sclass} but with a running Newton's constant $G_k$,
\be \label{Squant}
\Gb_k[g] = -Z_k \int d^d x\, \om\, R \, ,
\ee
where we define $Z_k = (16\pi G_k)^{-1}$. 
The gauge-fixing and ghosts will be discussed in the following.

\subsection{Variations}

Using \eqref{level-pert} we can obtain the general second variation of the action (up to boundary terms) as
\be
\begin{split}
\bigg(\f{\p^2}{\p\eps^2}   \int d^d x \,\om \,  R[g] & \bigg){\Big|_{\eps=0}} \\
  = \int d^d x \,\om  \bigg( & \f12  h^{(1)}{}_{\m\n} \nab^2 h^{(1)}{}^{\m\n} + (\nab^\m h^{(1)}{}_{\m\r}) (\nab^\n h^{(1)}{}^\r_\n) \\ 
 & + h^{(1)}{}_{\m\n} h^{(1)}{}_{\r\s} \Rb^{\m\r\n\s} + (h^{(1)}{}_{\m\r} h^{(1)}{}^\r_\n -h^{(2)}{}_{\m\n})\Rb^{\m\n}
 \bigg) \, .
\end{split}
\ee
In the exponential parametrization \eqref{exp-par}, $h^{(1)}{}_{\m\r} h^{(1)}{}^\r_\n -h^{(2)}{}_{\m\n}=0$ and the last term is therefore identically absent.
The same term is also absent for a generic parametrization on an Einstein background (i.e. satisfying $\Rb_{\m\n} = \f1d \gb_{\m\n}\Rb$) because of \eqref{unim2}.
Therefore, we conclude that on an Einstein background the Hessian is independent of the parametrization. 
This had to be expected, since the Einstein condition is precisely the equation of motion \eqref{eom}, and the on-shell Hessian is always independent of field redefinitions.
We see therefore that the exponential parametrization automatically gets rid of terms proportional to the field equations, but the logic we will follow here is the opposite: we will choose the background to be on-shell in order to obtain parametrization independence.

\subsection{Gauge fixing and TT decomposition}

The infinitesimal  gauge invariance of unimodular gravity is
\be
g_{\m\n}\to g_{\m\n} + \cL_v g_{\m\n} = g_{\m\n} + \na_\m v_\n +\na_n v_\m \, ,
\ee
where $\cL_v$ is the Lie derivative along a vector $v^\m$, which needs to preserve the unimodularity constraint, i.e. it must satisfy $g^{\m\n} \cL_v g_{\m\n}=\na_\m v^\m=0$. As a consequence of the latter, the vector  $v^\m$ has only three degrees of freedom.
Note also that as a consequence of unimodularity, the vector $v^\m$  must be a transverse vector both with respect to the full metric and to the background one, as can be seen by writing $\na_\m v^\m = \f{1}{\sqrt{g}} \,\p_\m (\sqrt{g} v^\m) =\f{1}{\sqrt{\gb}} \p_\m (\sqrt{\gb}\, v^\m) =\nab_\m v^\m$.

We fix the gauge as in \cite{Eichhorn:2015bna} (but keeping $\a$ generic), by including in the truncation the general gauge-fixing action
\be \label{gf}
\G_{\rm gf}[h, \gb] = \f{Z_k}{2\a} \int d^d x\sqrt{\gb} \hat{F}_\m \hat{F}^\m \, ,
\ee
with
\be
\hat{F}_\m \equiv [\Pi_{\rm T}]_\m^\a F_\a = (\d_\m^\a - \nab_\m\f{1}{\nab^2}\nab^\a) \nab^\n h_{\a\n} \, ,
\ee
$F_\a = \nab^\n h_{\a\n}$ being the usual de Donder gauge for the case $h_\m^\m=0$, and with the transverse projector $[\Pi_{\rm T}]_\m^\a = (\d_\m^\a - \nab_\m\f{1}{\nab^2}\nab^\a)$ (see e.g. \cite{BGMS}).
Note that the gauge-fixing action is already quadratic in $h_{\m\n}$ and therefore field redefinitions such as \eqref{redef} will not affect the FRG equation when we project this on the background metric, which we are allowed to do in the single-metric approximation.

Alternatively, we can project $\na^\n h_{\a\n}$ on its transverse part by a local (but higher-derivative) operator \cite{Alvarez:2008zw}:
\be \label{gf2}
\tilde{F}_\m \equiv  (\nab_\m\nab^\a -\nab^2 \d_\m^\a +\Rb_\m^\a) \nab^\n h_{\a\n}   \, .
\ee
It can be easily checked that $\nab^\m \tilde{F}_\m=0$; this is a consequence of $\tilde{F}_\m= - (-\nab^2 \d_\m^\a +\Rb_\m^\a)\hat{F}_\a $ and the fact that $\nab^\m(\D_{L,1})_\m^\a A_\a = -\nab^2 \nab^\m A_\m$ for any vector $A_\m$, and where $(\D_{L,1})_\m^\a \equiv (-\nab^2 \d_\m^\a +\Rb_\m^\a)$ is the Lichnerowicz Laplacian on one-forms. Note that in \cite{Eichhorn:2013xr} the Ricci term in  \eqref{gf} was missed.

Yet another way of imposing the same gauge condition is to replace $F_\m$ by its exterior derivative \cite{Buchmuller:1988yn}:
\be
\chi_{\m\n} = \p_\m F_\n - \p_n F_\m \, .
\ee
In fact we have (under integral and applying integrations by parts): $\chi_{\m\n}\chi^{\m\n} = 2 F^\m \tilde{F}_\m =  2 \hat{F}^\m \tilde{F}_\m$, where in the last step we used $\tilde{F}_\m = [\Pi_{\rm T}]_\m^\a \tilde{F}_\a$ and integrated by parts.

Ultimately all these choices of gauge-fixing functional (classically equivalent, but containing different orders of derivatives in $\G_{\rm gf}$) are completely equivalent once we take ghosts into account. In fact, as we will see, the gauge-fixing and ghost action completely cancel each other on an Einstein background, to which we will restrict from now on.

For a generic gauge, in order to reduce the Hessian to minimal form, we decompose the fluctuation field in transverse-traceless (TT) components (see e.g. \cite{BGMS}). Remembering that the trace $h_\m^\m=0$, the TT decomposition takes the form
\be \label{TT-dec}
h_{\mu\nu} = h_{\mu\nu}^{T} + \nab_\m \xi_\n + \nab_\n \xi_\m + \nab_\mu \nab_\nu \s - \frac{1}{d} \gb_{\mu\nu} \nab^2 \s \, ,
\ee
with the component fields satisfying
\be
\gb^{\mu \nu} \, h_{\mu\nu}^{T} = 0 \, , \quad \nab^\mu h_{\mu\nu}^{T} = 0
\, , \quad \nab^\mu \xi_\mu = 0 \, .
\ee
Then we have
\be
F_\m = \Big(\nab^2 +\f{\Rb}{d}\Big) \xi_\m +\nab_\m \Big(\f{d-1}{d}\nab^2 +\f{\Rb}{d}\Big)\s \, ,
\ee
and
\be
\hat{F}_\m  = \Big(\nab^2 +\f{\Rb}{d}\Big) \xi_\m  \, ,
\ee
rendering manifest the transverse nature of the gauge-fixing function. In fact, we might even directly choose $\bar{F}_\m=\xi_\m$ as gauge-fixing functional, as in \cite{Percacci:2015wwa}.

In order to decompose also the quadratic part of the action we need
%
%
\be
h^{\mu\nu}\Delta_{2} h_{\mu\nu} = \, h^{{\rm T}\,\mu\nu}\Delta_{2} h_{\mu\nu}^{\rm T} + 2\xi^\mu \Delta_{1}^2 \xi_\mu -  \sigma \left(\nab^2 + \tfrac{2\Rb}{d} \right)\left(\tfrac{d-1}{d}\nab^2 + \tfrac{\Rb}{d}\right)\nab^2 \sigma \, ,
\ee
where we defined
\be
\Delta_{2} h_{\mu\nu}  \equiv -\nab^2 h_{\mu\nu} - 2 \Rb_{\mu\,\,\,\nu}^{\,\,\,\alpha\,\,\,\beta} h_{\alpha\beta} \, ,
\ee
\be
\D_1 \xi_\mu \equiv -\Big(\nab^2 +\f{\Rb}{d}\Big) \xi_\m \, . 
\ee

Lastly, choosing \eqref{gf}, the quadratic part of the gauge-fixed gravitational action can be decomposed as
\be \label{hessian}
\begin{split}
\Big(  \f12 h^{\m\n}\cdot \Gb^{(2)}_{\m\n,\r\s}[\gb] \cdot h^{\r\s} + \G_{\rm gf}[h,\gb]   & \Big)_{|_{\rm Einstein\; space}}   = \\
   = \f{Z_k}{2} \int d^d x \,\om  \bigg( & \f12  h_{\m\n} \D_2 h^{\m\n} - F^\m F_\m +\f{1}{\a}  \hat{F}_\m \hat{F}^\m  \bigg) \\
   = \f{Z_k}{2}  \int d^d x \,\om \bigg( & \f12 h^{T\m\n} \D_2 h^T_{\m\n} 
    +\f{1}{\a} \xi^\m \D_1^2 \xi_\m \\
     & + \f{(d-1)(d-2)}{2d^2} \s' \nab^2 \s' \bigg) \, ,
\end{split}
\ee
where we defined
\be \label{sigma-prime}
\s' = \sqrt{-\nab^2} \sqrt{-\nab^2 - \f{\Rb}{d-1}} \s \, .
\ee
%

\subsection{Ghosts and auxiliary fields}

We follow \cite{Benedetti:2011ct} for the ghost sector, which reads
\be
S_{\rm gh} = \f{Z_k}{\a} \int d^d x \,\om \Big\{ \Cb^{T\,\m} \D_1^2 C_\m^T 
 + \f12  B^{T\, \m} \D_1^2 B_\m^T  \Big\}\, ,
\ee
with $C_\m^T$ and $B_\m^T$ two transverse vector fields, complex Grassmann and real, respectively. Note that bars on complex fields denote complex conjugation and there should be no confusion with the use of bars for quantities constructed out of the background metric.

The final ingredient of our truncation is the action for the auxiliary fields, introduced to take into account the Jacobians arising from the TT decomposition \eqref{TT-dec} and the redefinition \eqref{sigma-prime}:
\be \label{aux-gr}
S_{\rm aux} =  \int d^d x \,\om \Big\{   \bar{\chi}^{T\, \m} \D_1 \chi_\m^T   
 +  \z^{T\m} \D_1 \z_\m^T   \Big\}\, ,
\ee
where $\chi_\m^T$ and $\z_\m^T$ are two transverse vector fields, complex Grassmann and real, respectively.
Notice that the redefinition \eqref{sigma-prime} leads to the cancellation of the auxiliary scalar modes that would arise from the TT decomposition \cite{Dou:1997fg}.
We have no Jacobians for the ghosts, as the diffeomorphisms are transverse by construction.
%

\subsection{The FRG equation}

We use an adaptive cutoff $\cR_k$ such that $\D_2$, $\D_1$ and $\D_0\equiv -\nab^2$ get replaced in the total Hessian $\G_k^{(2)}+\cR_k$ by regularized operators, according to the rule $\D_s \to P_k(\D_s)= \D_s +R_k(\D_s)$, for the same cutoff function $R_k(x)$. 
Within such scheme, the FRG equation is
\be \label{frge}
\begin{split}
\p_t \G_k = & \f12 S\Tr \left[ \f{\p_t \cR_k}{\G_k^{(2)}+\cR_k} \right]  \\
= & \f12 \Tr_2 \left[ \f{\p_t R_k(\D_2)+\eta_k R_k(\D_2)}{P_k(\D_2)} \right] - \f12 \Tr_1 \left[ \f{\p_t R_k(\D_1)}{P_k(\D_1)} \right] \\
& + \f12 \Tr_0 \left[ \f{\p_t R_k(\D_0)+\eta_k R_k(\D_0)}{P_k(\D_0)} \right]
\end{split}
\ee
where we defined $\eta_k = \p_t \ln Z_k$.

Ghosts and gauge-fixing term have canceled exactly, and they do so even if we allow $\a$ to depend on $k$ (this was the motivation behind the peculiar construction of the ghost sector in \cite{Benedetti:2011ct}). Therefore, the flow is explicitly gauge-independent, as well as parametrization independent.

Note that if we were to follow the same scheme for the non-unimodular case, and view \eqref{unim} as a gauge fixing condition, we would have an additional scalar ghost arising from the Faddeev-Popov determinant $\sqrt{\det(\D_0)}$ associated to such a gauge \cite{Percacci:2015wwa}. This would contribute to the right-hand-side of \eqref{frge} with
\be \label{miss-ghost}
 -\f12 \Tr_0 \left[ \f{\p_t R_k(\D_0)}{P_k(\D_0)} \right] \, .
\ee
In such case, in the one loop approximation ($\eta_k=0$ on the right-hand-side), we recover the standard result, in which we are left only with the the spin 2 and spin 1 modes.

\subsection{Heat kernel}

The heat kernel expansion for an operator $\D_s$ provides an expansion in invariants of the form
\be\label{HKexp}
\Tr\left[ e^{- t \Delta_s} \right] = \left( \frac{1}{4 \pi t} \right)^{d/2} \int d^dx \,\om
\left\{ \tr a^{(s)}_0 + t \, \tr a^{(s)}_2 + t^2 \, \tr a^{(s)}_4 + \ldots \right\} \, ,
\ee
where $\tr$ denotes a trace with respect to the tensorial indices (for $s=1,2$).
The operators $\D_2$, $\D_1$ and $\D_0$ are the same used in \cite{BMS2}, and their associated heat kernel coefficients $a_i$ were derived there, although reported only for $d=4$. 
Here we need them for generic $d$, but only up to order $R$:
\begin{align}
& a^{(0)}_0 = 1 \, ,  & & a^{(0)}_2 = \frac{1}{6} \Rb \, ,\\
& \tr a^{(1)}_0 = d-1 \, ,  & &
\tr a^{(1)}_2 = \frac{d^2+5d-12}{6d} \Rb \, , \\
& \tr a^{(2)}_0 = \frac{1}{2} (d-2)(d+1) \, ,  & &
\tr a^{(2)}_2 = \frac{1}{12d} (d-6)(d+1)(d+4) \Rb \, .
\end{align}
In fact we do not need the $a^{(s)}_0$ coefficients, since in the unimodular case they add only a non-dynamical constant to the action. They correspond to standard vacuum terms, and in the unimodular case they are decoupled from any dynamics.

By use of a Mellin transform one can write the trace of a generic functional of $\D_s$ as \cite{Reuter:1996cp,Codello:2008vh}
\be
\Tr\left[W(\D_s)\right] = \f{1}{(4\pi)^{d/2}} \int d^d x\, \om \, \left( Q_{\f{d}{2}}[W] \,\tr a^{(s)}_0 + Q_{\f{d}{2}-1}[W]\, \tr a^{(s)}_2 + \ldots  \right) \, ,
\ee
where for $n>0$ we have
\be
Q_{n}[W] = \f{1}{\G(n)} \int_0^\infty dz\, z^{n-1} W(z) \, .
\ee
We do not need $n\leq 0$ since we will be concerned only with $d>2$ and no higher derivative terms.

Writing $R_k(\D) = k^2\, r (\D/k^2)$, and defining $z=\D/k^2$, we find that we need to evaluate the following two integrals:
\be
I_d = \f{1}{\G(\f{d}{2}-1)} \int_0^\infty dz\, z^{\f{d}{2}-2} \f{r(z)-z r'(z)}{z+r(z)} \, ,
\ee
\be
J_d = \f{1}{\G(\f{d}{2}-1)} \int_0^\infty dz\, z^{\f{d}{2}-2} \f{r(z)}{z+r(z)} \, .
\ee

For $r(z)=(1-z)\th(1-z)$ (also known as ``optimized cutoff'' \cite{Litim:2001up}) we find
\be
I^{\rm opt}_d = \f{2}{(d-2)\G(d/2-1)} \, , \;\;\; J^{\rm opt}_d = \f{4}{d (d-2)\G(d/2-1)} \, ,
\ee
while for $r(z)=z/(e^z-1)$ (also known as ``exponential cutoff'') we find
\be
I^{\rm exp}_d = \f{d-2}{2} \z(d/2) \, , \;\;\; J^{\rm exp}_d = 1 \, .
\ee
%

\subsection{Beta function and fixed point}

Using \eqref{frge}, projected on the background (i.e. taking $h_{\m\n}=0$) and \eqref{HKexp} we obtain
\be \label{frge-R}
- \p_t Z_k \int d^d x\, \om\, \Rb =
 \f{1}{(4\pi)^{d/2}} \int d^d x\, \om \, \left( I_d\, \Big(\tr a^{(2)}_2 -\tr a^{(1)}_2+a^{(0)}_2\Big) + J_d \f{\eta_k}{2} \Big(\tr a^{(2)}_2 +a^{(0)}_2\Big) \right) \, ,
\ee
or
\be
- \p_t Z_k  = \f{k^2}{(4\pi)^{d/2}}  \left( I_d\, \Big(\f{d^2-3d-34}{12}\Big) 
+ J_d \f{\p_t Z_k}{Z_k} \Big(\f{d^2-d-24}{24}-\f{1}{d}\Big) \right) \, ,
\ee
that is,
\be
\p_t \tG = \b(\tG)\, ,
\ee
\be
\b(\tG)\equiv (d-2) \tG + \tG^2 \f{ \Big(\f{d^2-3d-34}{12}\Big) I_d}{\pi(4\pi)^{\f{d}{2}-2} +  \Big(\f{d^2-d-24}{24}-\f{1}{d}\Big)  J_d\, \tG}\, ,
\ee
where we introduced the dimensionless Newton's constant $\tG=k^{d-2}G_k$.

Note once more the explicit gauge independence of the beta function, which we obtained by working on shell. In fact, the final result does not depend on the choice of background, as it is easy to understand by the following argument: suppose we had worked off-shell and found additional terms in \eqref{frge-R}; we would still be able to impose the field equations in such final result, and we should obtain again \eqref{frge-R}; therefore, the additional terms should be proportional to the equations of motion, i.e. they should have the form $X_{\m\n} (\Rb^{\m\n} -\f1d \gb^{\m\n}\Rb)$, but since the only tensor that can appear in $X_{\m\n}$ at this order of the heat kernel expansion is $\gb_{\m\n}$ we find that such terms must vanish due to the tracelessness of the field equations.

Having obtained the beta function, we can turn our attention to its fixed points, which of course will also be  gauge and parametrization independent.
By definition, a fixed point $\tG^*$ satisfies the equation $\b(\tG^*)=0$. Besides the GFP, we find the non-trivial fixed point
\be \label{NGFP}
\tG^* = -\frac{3\times 2^{d-1} (d-2) d \pi ^{\frac{d}{2}-1}}{2 d \left(d^2-3 d-34\right)
   I_d+\left(d^4-3 d^3-22 d^2+24 d+48\right) J_d}\, .
\ee

Both with the optimized and the exponential cutoff, $\tG^*$ is positive in a range $d\in(2,d_c)$ and it satisfies the limits $\lim_{d\to 2} \tG^*=0$ and $\lim_{d\to d_c^-} \tG^*=+\infty$. In order to understand the cutoff dependence of $d_c$ we note that since $r(z)\geq 0$ and $r'(z)\leq 0$ (where of course for no valid cutoff the equal sign is realized at all $z$) we have $I_d>0$ and $J_d>0$.
It is then clear that the denominator of \eqref{NGFP} can become zero only if the coefficients in front of these two functions have a different sign, which happens for $5.82\lesssim d \lesssim 7.52$. For $d \lesssim 5.82$ both coefficients are negative, making $\tG^*>0$.
We thus conclude that within our truncation, independently of the gauge, of the parametrization, and of the cutoff, unimodular gravity in $d=4$ has a non-trivial fixed point.
A rough estimate of the fixed-point value in $d=4$ for generic cutoff is obtained by noting that both $I_d$ and $J_d$ are of order one (again a consequence of general cutoff properties), and therefore
\be
\tG^*{\big|_{d=4}} = \frac{4 \pi }{5 I_4+3 J_4} \sim 1\, .
\ee
In particular, for the optimized cutoff we find $\tG^*{\big|_{d=4}}=8\pi/13\simeq 1.93$, and for the exponential one we find $\tG^*{\big|_{d=4}}=24\pi/(18+5\pi^2) \simeq 1.12$.

The critical exponent associated to the fixed point is
\be
\f{1}{\n} = - \f{\p \b(\tG)}{\p \tG}\Big|_{\tG=\tG^*} = \frac{(d-2) \left(2 d \left(d^2-3 d-34\right) I_d+\left(d^4-3 d^3-22 d^2+24
   d+48\right) J_d\right)}{2 d \left(d^2-3 d-34\right) I_d} \, ,
\ee
which in $d=4$ becomes
\be
\f{1}{\n}\Big|_{d=4}  = 2+ \f{6 J_4}{5I_4} \, ,
\ee
and by the same reasoning as above we can say that $\n|_{d=4}\sim 1/3$.
For the optimized cutoff we find $\n^{-1}=13/5 = 2.6$, and for the exponential one we find  $\n^{-1}=2+36/(5\pi^2) \simeq 2.73$.

To conclude, we notice that in $2+\eps$ dimensions we have (in the one-loop approximation and for infinitesimal $\eps$)
\be
\b(\tG) = \eps \tG - 12\, \tG^2 \, ,
\ee
independently of the choice of cutoff.
Rewriting $12= \f23 (18-c)$ with $c=0$ ($c$ being the central charge of matter fields), we observe that the discrepancy with the result obtained in the non-unimodular case, i.e. $\f23 19$ (obtained in the unimodular gauge in \cite{Percacci:2015wwa}) is due precisely to the missing ghost \eqref{miss-ghost} (having $c=-1$).

\section{Conclusions}
\label{concl}

In this paper we have pointed out the different nature of Newton's constant in unimodular gravity as compared to the non-unimodular case. While in the latter $G$ can be absorbed by a rescaling of the metric, this is not possible in UG, where therefore Newton's constant is an essential coupling.
We have illustrated the consequent gauge and parametrization independence of its flow via a simple approximation.
We conclude here with two more comments.
 
First, one should  notice that even though $G$ is an essential coupling in UG, this still does not mean that its running can be directly translated in a physical prediction: at the perturbative level, $G$ gets renormalized by quadratic divergences, and this means that the running is not universal, as discussed in \cite{Anber:2011ut}. And in fact we found that the beta function still depends (even at one-loop, with the exception of $2+\eps$ dimensions) on the choice of cutoff. Non-universality of the running is however not a problem for us, as in a Wilsonian RG flow everything runs, and the purpose is to define the theory in terms of relevant perturbations of a fixed point. Extracting the physical scale-dependence of scattering amplitudes and other observables is a separate question. On the contrary, even in a Wilsonian flow it is important to distinguish redundant (or inessential) couplings, as these are generally not required to have a fixed point. Therefore, we think that UG presents a clear advantage from this point of view.

We should also stress once more the obvious fact that most of our results and reasoning carry over to the non-unimodular case once we use the unimodular constraint as a (partial) gauge fixing.
As a consequence, if one is convinced by our argument that unimodularity is a favorable option from the RG point of view, there still remains open the question of whether we should think of that as a fundamental part of the definition of the theory, or just as a favorable gauge fixing choice. We tend to think that the first option is more natural, because the notion of a preferred gauge fixing is in contrast to standard wisdom on the unphysical role of gauge fixing.

Although the question above might appear at first philosophical in nature, we should stress again that in fact the difference between fundamental unimodularity and unimodular gauge has practical consequences at a numerical level. The absence of the ghost \eqref{miss-ghost} leads to slightly different results at a quantitative level. In particular, one would probably find a difference in the quantum corrections to Newton's potential (computed in detail in the non-unimodular case in \cite{BjerrumBohr:2002kt}), which in principle would be observable. We hope to explicitly check this in the near future.


\providecommand{\href}[2]{#2}\begingroup\raggedright\endgroup

\end{document}